\documentclass[10pt,letterpaper]{article}
\usepackage{amsmath}
\usepackage[top=0.85in,left=0.75in,footskip=0.75in,marginparwidth=2in]{geometry}

\usepackage[utf8]{inputenc}

\usepackage{cite}

\usepackage{nameref,hyperref}

\usepackage[right]{lineno}

\usepackage{microtype}
\DisableLigatures[f]{encoding = *, family = * }

\raggedright
\usepackage{changepage}

\usepackage[aboveskip=1pt,labelfont=bf,labelsep=period,singlelinecheck=off]{caption}

\makeatletter
\renewcommand{\@biblabel}[1]{\quad#1.}
\makeatother

\usepackage{lastpage,fancyhdr,graphicx}
\usepackage{epstopdf}
\fancyheadoffset[L]{2.25in}
\fancyfootoffset[L]{2.25in}

\usepackage{color}

\definecolor{Gray}{gray}{.25}

\usepackage{graphicx}

\usepackage{sidecap}

\usepackage{wrapfig}
\usepackage[pscoord]{eso-pic}
\usepackage[fulladjust]{marginnote}
\reversemarginpar

\begin{document}
\vspace*{0.35in}

\begin{flushleft}
{\Large
\textbf\newline{Zero-Order Diffraction Suppression in Full Field-of-View Computer Generated Holography: A Camera In the Loop Interferometric Approach}
}
\newline
\\
Alessandro Cerioni\textsuperscript{1,*},
Samuele Trezzi\textsuperscript{1},
Marco Astarita\textsuperscript{1},
Tommaso Ongarello\textsuperscript{2},
Anna Cesaratto\textsuperscript{2},
Giulio Cerullo\textsuperscript{1},
Andrea Bassi\textsuperscript{1},
Gianluca Valentini\textsuperscript{1},
Paolo Pozzi\textsuperscript{1}
\\
\bigskip
\bf{1} Physics Department, Politecnico di Milano, Milan, Italy
\\
\bf{2} EssilorLuxottica Smart Eyewear Lab, via Pascoli 70/3, 20133 Milan, Italy
\\
\bigskip
* alessandro.cerioni@polimi.it

\end{flushleft}

\section*{Abstract}
We introduce a novel interferometric approach for suppressing zero-order diffraction (ZOD) in phase-only computer-generated holography. The technique relies on the destructive interference between the zeroth-order light and a suppression beam in a plane optically conjugated to the spatial light modulator (SLM). A camera-in-the-loop (CITL) calibration procedure retrieves the optimal pixel-wise phase map that cancels out the ZOD component with high precision, while preserving the full modulation depth of the SLM. Experimental demonstrations on point-cloud and 2D/3D holograms achieve up to $99\%$ suppression of the ZOD intensity, without loss of image quality or field of view. Once calibrated, the correction can be applied to any hologram without recomputation, enabling real-time operation and robust performance over time. This method removes a long-standing barrier to the practical deployment of full-field holography, facilitating the development of compact, high-fidelity holographic engines for augmented and mixed reality displays.


\section{Introduction}
Phase-only computer generated holography (CGH) is employed in numerous applications, including beam shaping \cite{piccardo2023broadband}, data storage, 3D photostimulation for optogenetics \cite{pozzi2018fast}, optical manipulation \cite{grier2003revolution}, and 3D imaging for Augmented Reality (AR) \cite{gopakumar2024fullcolour,jang2024waveguide,Astarita:25}. Among these, 3D CGH for AR has gained significant relevance due to the growing interest in AR applications for near-eye displays (NEDs), driven by the numerous benefits offered by laser-based phase-only CGH.
The adoption of CGH for 3D virtual images in AR offers the advantage of eliminating the vergence-accommodation conflict (focus rivalry) \cite{pan2015review, VAC}, a major challenge currently faced by existing NED technologies. Additionally, thanks to new algorithms and devices, CGH can provide aberration-free, high-resolution images and enables real-time  videos\cite{CGHalg2022}. The use of laser-based CGH will also facilitate outdoor implementation for AR, due to the high brilliance of laser light, making it suitable for high-light conditions. Next-generation NEDs may achieve significant performance enhancements, directly leveraging the innovations introduced in CGH technology.

Although holographic technology is considered the ultimate 3D display approach \cite{jang2024waveguide} and can deliver realistic virtual images, many limitations still need to be overcome. These include not only the significant computational effort required to calculate holograms but also the limitations of devices that modulate light.

In particular, among other limitations, spatial light modulators (SLMs) used in holographic displays are pixelated devices and, as such, they cannot modulate all the light that impinges on them. Indeed, part of the light is reflected back specularly due to various causes, such as unmodulated inter-pixel regions and imperfect anti-reflection coatings, resulting in an unwanted bright spot at the center of any virtual image produced by the SLM, known as the zeroth order diffraction (ZOD). 
The presence of non-modulated light is a common issue even in advanced implementations \cite{jang2024waveguide} and is usually removed in various ways. The most common method involves spatial filtering by placing an obstructive target in the center of the Fourier plane (2f) of a 4f telescope positioned after the SLM, blocking the non-modulated reflected light at low spatial frequencies. However, this approach requires a bulky 4f system, incompatible with miniaturized AR devices, and results in a dark spot in the center of the virtual image.

Alternative approaches involve shifting the projected light away from the ZOD position: laterally, axially or both\cite{Yun:23,LIANG2022107048}. However, lateral displacement limits the effective field of view and reduces the diffraction efficiency, introducing comatic aberrations into the projected light pattern \cite{palima2007holographic}. The axial displacement causes the ZOD to shift significantly out of focus from the  computed image, but the imposed quadratic phase significantly reduces the axial field of view of the device.

Recent studies have demonstrated ZOD suppression through destructive interference schemes that combine two-photon excitation with aberration-based techniques \cite{hernandez2014ZOD,yu2024performance}. Although these nonlinear approaches have proven effective for multiphoton microscopy, their intrinsic non-linearity makes them incompatible with CGH in the visible wavelength range required by NEDs. In parallel, several linear strategies have been explored to mitigate the zero-order component directly at the SLM level. For example, phase-compression methods \cite{liang2012ZOD} reshape the phase profile to minimize residual non-modulated light, achieving experimentally only a factor of 3 intensity reduction (i.e., $~67\%$ suppression) with degradation in diffraction efficiency for varying compression factors, thus offering limited robustness across different hologram patterns. Similarly, interferometric correction schemes based on an auxiliary beam \cite{Improso2017SuppressionOZ} promise high suppression through destructive interference (up to $99\%$ in simulations), but experimentally deliver just $32\%$ due to mismatches in phase and amplitude calibration.

In this work, we introduce a camera-in-the-loop interferometric calibration that retrieves a system-specific, pixel-wise corrective phase map to suppress the ZOD via destructive interference. After calibration, the same phase map can be applied to arbitrary holograms, enabling full field of view operation without external spatial filtering.

\section{Method}
In Fourier-transform CGH, the above mentioned unwanted light component manifests as a strong ZOD peak in the center of the spatial-frequency plane, where the virtual image is generated. Our method engineers the SLM's phase pattern to shape a portion of the modulated light into a corrective optical field. This field is designed to have the same intensity and the exact opposite phase as the native ZOD, allowing the two fields to cancel each other upon interference and effectively suppress the central spot without degrading the reconstructed image.
The entire process is divided into four key stages: 
\begin{enumerate}
  \item Initial estimation of the ZOD corrective beam amplitude.
  \item Application of a uniform phase pattern (“piston sweep”) to probe the corrective beam phase.
  \item Pixel-resolved retrieval of the non-uniform corrective phase map using the camera.
  \item Precise mapping of the corrective phase from camera to SLM coordinates, followed by recursive optimization until convergence.
\end{enumerate}
A schematic of the iterative process is reported in Fig. \ref{fig:setup}(d).

The procedure requires a dedicated setup, shown in Fig. \ref{fig:setup}(a), that includes a camera (Cam1) conjugated to the SLM plane through a 4f telescope.  A pinhole is positioned in the Fourier plane of the first lens (2f), in order to block all the hologram light not directly addressed at the ZOD location, therefore, behaving as a low pass filter. A second camera (Cam2) images the Fourier plane without any spatial filtering.

This second camera is also employed for the precise calibration of the SLM to ensure a linear phase response over the entire $0$ to $2\pi$ range. An accurate phase calibration of the SLM, as detailed in \cite{ronzitti2012LCOS}, is a crucial prerequisite for the correct implementation of the proposed method. The four steps of the process are described in details below.\\
After retrieving the ideal phase map, the 4f setup can be dismantled and the SLM can be used with no additional optics, or in a separate optical setup.

\subsection{Coefficient Estimation of Interfering Beams}
The process starts by estimating the ratio between the amplitude of the ZOD field and that of the hologram. This is done by uploading a phase map $\Phi_{\mathrm{Holo}}$ to the SLM that encodes an arbitrary virtual image, with the unique constraint of contributing zero intensity at the ZOD location (i.e., the center of the Fourier plane). Under this condition, the optical field at the SLM plane in this uncorrected configuration is

\begin{equation}
E_{\mathrm{no\ corr}} = A\, e^{-i\Phi_{\mathrm{Holo}}} + B\, e^{-i\Phi_{\mathrm{ZOD}}}
\label{eq:no_corr}
\end{equation}

where $\Phi_{\mathrm{Holo}}$ is the phase pattern for the virtual image, $\Phi_{\mathrm{ZOD}}$ is the inherent phase of the native ZOD, $A$ is the amplitude of the modulated light associated with the virtual image, and $B$ is the amplitude of the native unmodulated ZOD component. It is critical to notice that, while $\Phi_{\mathrm{ZOD}}$ has been considered a scalar value in previous literature (e.g., \cite{yu2024performance}), in this method we consider it as a function of the spatial coordinates at the SLM plane, i.e. $\Phi_{\mathrm{ZOD}}(x,y)$.

In this uncorrected configuration, the total intensity $I_{\mathrm{tot}}$ (measured in arbitrary units by integrating over the entire detection area of Cam2) is

\begin{equation}
I_{\mathrm{tot}} = A^2 + B^2
\label{eq:balance_no_corr}
\end{equation}

with no interference term due to the design constraint of the virtual image. The ZOD intensity $I_{\mathrm{ZOD}}$, measured in a small region of interest (ROI) around the center of the Fourier plane (ZOD location) on Cam2, is then simply

\begin{equation}
I_{\mathrm{ZOD}} = B^2,
\end{equation}

as no modulated light contributes there. These measurements allow for direct estimation of $B = \sqrt{I_{\mathrm{ZOD}}}$ and, subsequently, $A^2 = I_{\mathrm{tot}} - B^2$.

The ratio of total to ZOD intensity from this uncorrected measurement provides the following relation:

\begin{equation}
\frac{I_{\mathrm{tot}}}{I_{\mathrm{ZOD}}} = \frac{A^2 + B^2}{B^2} = \frac{A^2}{B^2} + 1.
\label{eq:ratio}
\end{equation}

To introduce ZOD suppression, a corrective replica field must be added to the overall field. The total optical field at the SLM plane now becomes

\begin{equation}
E_{\mathrm{corr}} = A_1\, e^{-i\Phi_{\mathrm{Holo}}} + A_2\, e^{-i\Phi_{\mathrm{Corr}}} + B\, e^{-i\Phi_{\mathrm{ZOD}}}
\label{eq:corr}
\end{equation}

where $A_1$ is the adjusted amplitude for the virtual image term, $A_2$ is the amplitude of the corrective ZOD replica, and $\Phi_{\mathrm{Corr}}$ is its phase.

For destructive interference to cancel the ZOD, the corrective replica must match the native ZOD in amplitude and be opposite in phase:

\begin{equation}
A_2 = B, \quad \Phi_{\mathrm{Corr}} = \Phi_{\mathrm{ZOD}} + \pi
\end{equation}

Energy conservation requires that the total modulated intensity remains consistent with the uncorrected case. Since the corrective spot is placed in the same location as ZOD, there is no significant interference between correction beam and virtual image, leading to   

\begin{equation}
A^2 = A_1^2 + A_2^2
\label{eq:A_def}
\end{equation}

Substituting $A_2 = B$ into Eq.~\eqref{eq:A_def} and solving for $A_1$ using the measured ratio from Eq.~\eqref{eq:ratio} yields

\begin{equation}
A_1 = \sqrt{A^2 - A_2^2} = \sqrt{A^2 - B^2} = B \sqrt{\frac{I_{\mathrm{tot}}}{I_{\mathrm{ZOD}}} - 2}.
\label{eq:A1}
\end{equation}

This ensures that the virtual image retains sufficient energy, while allocating exactly $B^2$ to the corrective term. In practice, we additionally verify and refine this model-based weighting by performing a short 1D search around the estimated ratio $A_2/A_1$. We evaluate the residual ZOD peak on Cam2 for a few trial ratios and select the one that yields the minimum, which compensates for small measurement errors in $I_{\mathrm{ZOD}}$ and $I_{\mathrm{tot}}$ and ensures the best achievable suppression.
With the final $A_1$ and $A_2$ values determined, the initial SLM phase map is constructed as

\begin{equation}
\Phi_{\mathrm{SLM}} = \arg\left( A_1\, e^{-i\Phi_{\mathrm{Holo}}} + A_2\, e^{-i\Phi_{\mathrm{Corr}}} \right),
\label{eq:phaseslm}
\end{equation}

which initiates the ZOD suppression process. The last relation between the amplitude of the virtual image and that of the ZOD (Eq.~\eqref{eq:A1}) implies that full ZOD removal is possible only if the overall modulated intensity is at least twice that of the ZOD. This condition is easily met by modern phase-only SLMs.

\subsection{Uniform phase pattern application "pistons"}
\label{sec:pistons}
Once the amplitude coefficients $A_1$ and $A_2$ have been estimated, we perform the precise phase optimization step using a CITL method based on Cam1.
Since the corrective beam phase is initially unknown, we begin by applying a series of uniform-phase maps (“piston sweep”) as phases of the correction beam (see Eq. \ref{eq:phaseslm}). The resulting interference intensity between the correction field and the native ZOD field is recorded for every applied uniform phase value on the Cam1 in Fig. \ref{fig:setup}(a). A pinhole is placed at the Fourier plane of the 4f telescope, acting as a spatial filter that transmits only the ZOD and the corrective beam (C) while blocking the higher-frequency components of the virtual image. The pinhole is chosen with a wider diameter than the systems's diffraction limit, in order ot account for aberrations in the lens between the SLM and the pinhole plane. The minimum resolvable piston-phase step is set by the calibrated $0 - 2\pi$ modulation range of the SLM, divided by the number of available phase quantization levels.

Figure \ref{fig:setup}(b) displays the intensity response of three sample pixels, each following a cosine dependence on the applied piston phase, with minima occurring at different piston values.

\subsection{Non-Uniform Phase Retrieval}
\label{sec:nonuniform}
The diversity of the phase values that must be applied to minimise the intensity in different pixels clearly indicates that the ideal phase to be applied to the corrective beam is inherently non-uniform across the SLM surface. This non-uniformity accounts for any inherent phase variations in the native ZOD beam itself.

To identify the specific phase $\phi_{i,{\min}}$ that yields the minimum ZOD intensity for each camera pixel $(x_{cam},y_{cam})$, we applied the piston values for all quantization levels $\phi_i$ (where $i=1,\dots,N$ refer to the distinct piston phases) and recorded the corresponding 2D intensity frames $I_i(x_{cam},y_{cam})$. By stacking these $N$ frames, we construct a three-dimensional \emph{Intensity array} $\mathcal{I_M}(x_{cam},y_{cam},i)$, where the first two dimensions refer to the coordinates of each camera pixel, and the third dimension refers to the applied piston phase. For each pixel we can select the phase value minimizing the intensity by:
\begin{equation}
\Phi_{\mathrm{optimized}}(x_{\mathrm{cam}},y_{\mathrm{cam}})
=
\arg\min_{i}\,\mathcal{I}_M(x_{\mathrm{cam}},y_{\mathrm{cam}},i)\,.
\end{equation}

To further enhance the accuracy and reduce measurement noise, each intensity slice $I_i$ is initially smoothed with a Gaussian filter before determining its minimum. This further phase optimization step ultimately delivers a high-resolution Phase map $\Phi_{\mathrm{cam}}(x_{cam},y_{cam})$, which contains the best phase values. Such spatially-resolved phase map is then fed into the subsequent affine-mapping step (Section~\ref{sec:affine}) to generate the final SLM phase pattern.

\subsection{Affine Transformation}
\label{sec:affine}
Due to inherent disparities in size, pixel pitch, and orientation between the camera sensor and the SLM panel, establishing an accurate spatial correlation of plane coordinates is critical. This is achieved through an affine transformation that maps the optimized phase map $\Phi_{\mathrm{cam}}(x_{\mathrm{cam}}, y_{\mathrm{cam}})$, computed in camera coordinates, onto the SLM's physical grid $(x_{\mathrm{slm}}, y_{\mathrm{slm}})$. The transformation is governed by an affine matrix $2 \times 3$ $\mathbf{M}$, which encompasses linear geometric transformations such as scaling, rotation, shear, and translation, and is expressed as

\begin{equation}
\begin{pmatrix}
x_{\mathrm{cam}} \\
y_{\mathrm{cam}}
\end{pmatrix}
=
\mathbf{M}
\begin{pmatrix}
x_{\mathrm{slm}} \\
y_{\mathrm{slm}} \\
1
\end{pmatrix}.
\end{equation}

To determine $\mathbf{M}$, a series of reference marker patterns are sequentially projected onto the SLM, and their corresponding positions $(x_{\mathrm{cam},i}, y_{\mathrm{cam},i})$ are recorded by Cam1. With a minimum of three non-collinear point correspondences, the affine matrix can be uniquely computed through least square minimization method.

\begin{figure*}[htp]
\centering
\includegraphics[width=\textwidth]{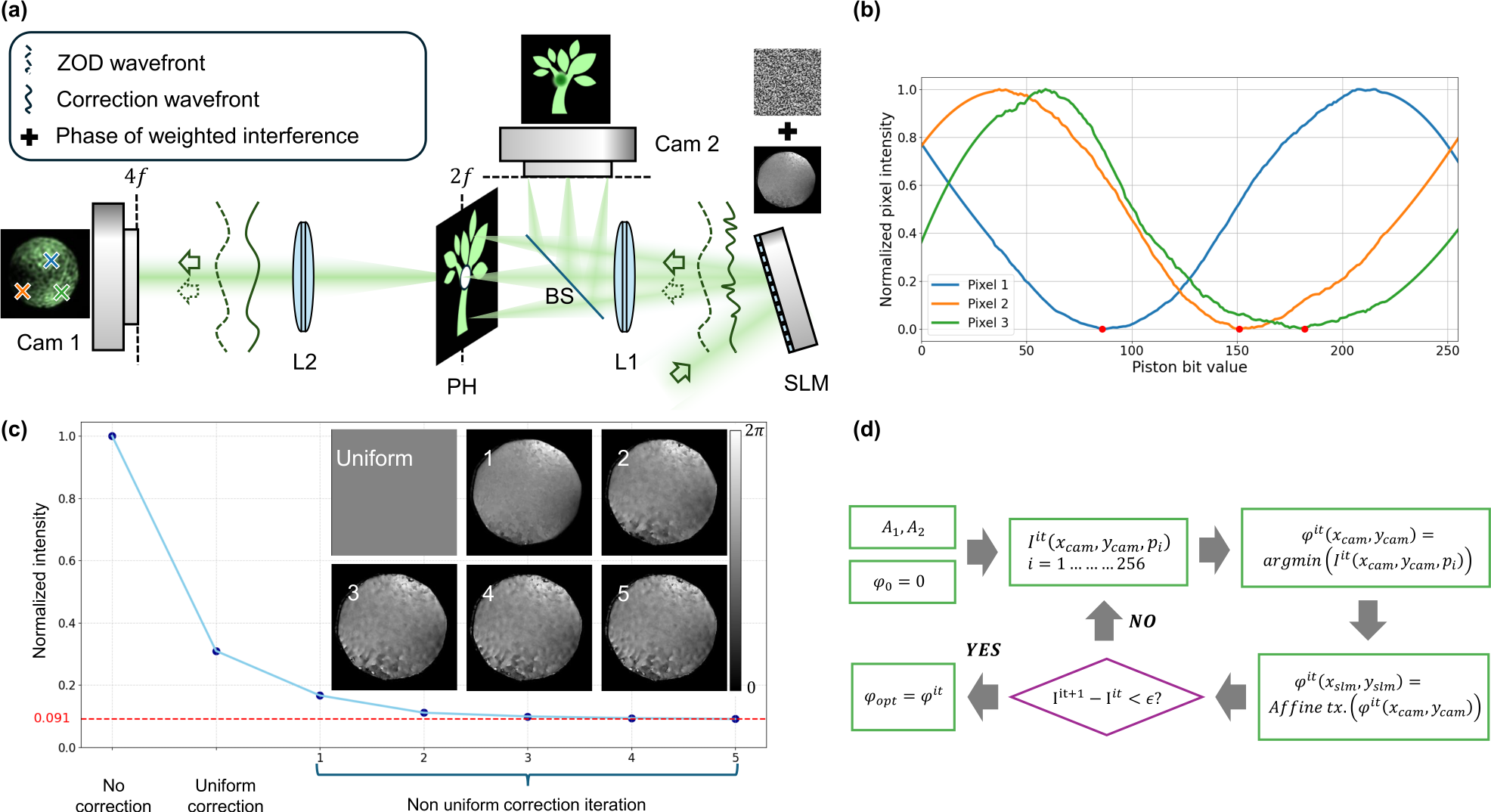}
\caption{(a) Schematic of the 4f setup with unitary magnification. A 300 $\mu$m pinhole placed at the Fourier plane acts as a low-pass filter, allowing only the zero-order diffraction (ZOD) and the corrective beam to pass. The camera records the spatial interference between these two beams. (b) The recorded intensity for three representative pixel follow the expected cosinusoidal dependence on the applied piston phase. The minima occur for different values of the phase shift, therefore, allowing for an optimal bidimensional corrective phase map. (c) The integral of the pixel intensity over the entire camera sensor in the 4f configuration decreases with each iteration and quickly converges, with negligible improvement after only a few cycles. The retrieved phase maps at successive iterations are also shown, highlighting that the updates become progressively smaller. (d) Schematic representation of the employed algorithm for ideal phase retrivial.}
\label{fig:setup}
\end{figure*}

Following calibration, the derived matrix $\mathbf{M}$ is applied to transform any optimized phase pattern. Given that phase values are discrete and periodic (with $0$ and $2\pi$ representing equivalent phase shifts), averaging interpolations (e.g., bilinear or bicubic) must be avoided: such methods would incorrectly average values across the $0/2\pi$ discontinuity, yielding an erroneous phase near $\pi$ (a $180^\circ$ shift) that deviates substantially from the intended physical phase. Instead, nearest-neighbor interpolation is enforced to preserve the exact, physically meaningful phase values without introducing artifacts. 
The resulting output array, \texttt{phase\_SLM}, represents the final phase pattern accurately mapped to the SLM's native resolution, ensuring that each correction-phase value aligns with the intended physical pixel for accurate ZOD cancellation. 
While from a purely theoretical standpoint a full correction should be achieved with a single measurement, experimental errors and aberrations in the measurement setup will affect the quality of the estimate. In order to improve the correction, the retrieved non-uniform phase map then serves as the new baseline for further "Piston" based iterations, as explained in Section \ref{sec:pistons} of the optimization process.
With each subsequent iteration, the discrepancy between the target ideal phase (which induces destructive interference) and the retrieved phase is progressively reduced. By repeatedly updating the optimal parameters using the latest intensity matrix, errors and discrepancies from the ideal wavefront are steadily mitigated. This results in a consistent decline in ZOD intensity at every stage, yielding better destructive interference. This iterative procedure is repeated until the intensity of the two interfering beams ceases to decrease further, as determined by the difference in intensity measured on Cam1 across consecutive assessments falling below a specified threshold. The characteristic convergence pattern, showing the drop in ZOD intensity versus iteration count, is depicted in Fig. \ref{fig:setup}(c).

The experimental setup employed to evaluate the closed-loop ZOD suppression method, as depicted in Fig. \ref{fig:setup}(a), is based on two lenses of focal length $f = 100 \ \text{mm}$ (AC508-100-A-ML, Thorlabs, USA) that make a 4f system with unity magnification. The light source is a single-mode laser (C-FLEX, HÜBNER Photonics, Germany) emitting at 532 nm,linearly polarized (horizontal polarization as requested by the SLM), collimated by a lens (AC254-050-A-ML, Thorlabs, USA) and cropped with an iris (IDA25/M, Thorlabs, USA) to match the SLM dimensions. The SLM used in all experiments is the Holoeye PLUTO-2.1, a phase-only reflective LCoS panel with 1920 × 1080 resolution, 8.0 µm pixel pitch, 93\% fill factor, and 15.36 × 8.64 mm active area. The beam splitter (BS, BSW27, Thorlabs, USA) splits the modulated light into two paths: one directed to a secondary camera (Cam2) (Panda 4.2, PCO, DE) at the 2f plane (Fourier domain) to capture the ZOD as a bright spot at zero spatial frequency, and the other passing through a 300 $\mu$m pinhole at the Fourier plane (2f). The pinhole acts as a low-pass filter and isolates the ZOD and the correction beam, that are imaged by the second lens to the primary camera (Cam1) (Panda 4.2, PCO, USA) at the 4f plane, which is optically conjugated to the SLM. The primary camera records the interference between the ZOD and correction beam, enabling precise measurement of their phase relationship for iterative optimization. The secondary camera at the 2f plane provides simultaneous monitoring of the Fourier domain, facilitating a comprehensive analysis of the ZOD suppression process.
All calculations reported in the Method section were performed through custom Python code, using the OpenCV library for affine transform estimation and application.

\section{Results}

Experimental validation of our CITL interferometric method demonstrates robust suppression of the ZOD artifact across multiple holographic configurations, highlighting both its versatility and its relevance for high-quality augmented reality imaging. To enable quantitative assessment, we first considered a 2D point-cloud hologram and a Fourier-transform hologram of a planar tree structure, where the total hologram intensity can be precisely measured on the 2D camera sensor.

\begin{figure*}[htp]
\centering
\includegraphics[width=\textwidth]{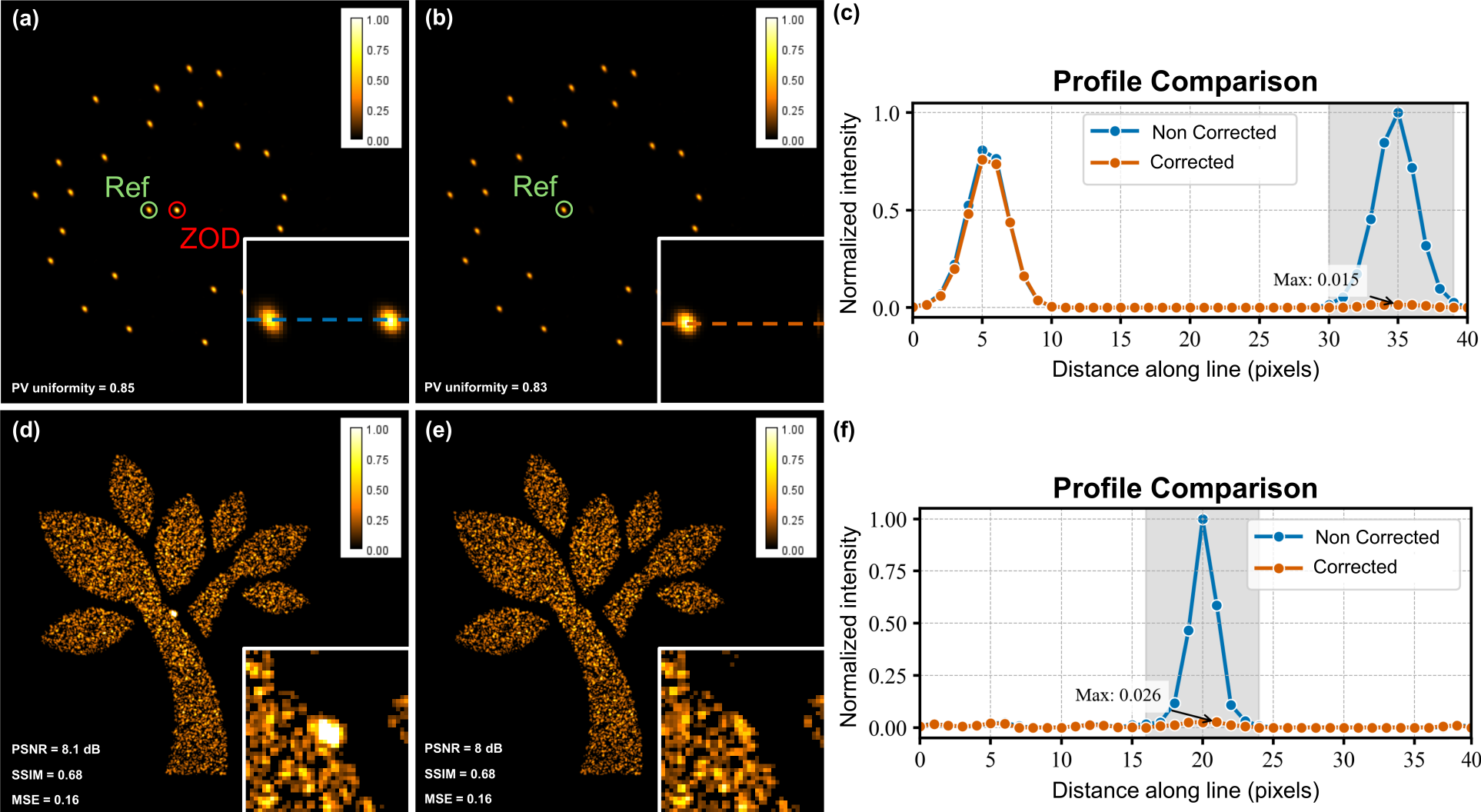}
\caption{Experimental demonstration of ZOD suppression. (a) Reconstructed point-cloud hologram without correction and (b) with correction. The magnified view of the ZOD central region in (a) shows a bright ZOD peak, while the corresponding view in (b) shows its elimination. (c) Quantitative intensity profile extracted from the magnified insets of (a) and (b), comparing the uncorrected (blue dashed line) and corrected (orange dashed line) cases. (d) Reconstructed tree hologram without correction and (e) with correction. The magnified box in (d) shows the ZOD, which is removed in (e). For visibility reasons ZOD of the uncorrected case is intentionally allowed to saturate: both figures are normalized to the maximum value of the uncorrected holographic image (d), excluding the ZOD region. (f) Quantitative intensity comparison for the tree hologram's central region for non saturating image. Image quality metrics (PV uniformity, PSNR, SSIM, MSE) are nearly identical for the uncorrected and corrected holograms. These results confirm that the proposed correction consistently suppresses the ZOD without degrading the holographic image quality.}
\label{fig:holograms}
\end{figure*}

Fig.\ref{fig:holograms} from (a) to (e) show a quantitative comparison for the point-cloud hologram: the uncorrected point-cloud image (a) exhibits a ZOD artifact that has similar brightness to the points of the cloud. Such pronounced disparity highlights the critical obstacle to successfully integrating a holographic-based system into an AR NED architecture without implementing a robust, active ZOD suppression mechanism.

However, the application of our corrective technique yields a significant enhancement, as demonstrated in the corrected image of Fig. \ref{fig:holograms}(c). Specifically, the system achieves a decisive reduction in ZOD artifact irradiance, quantified at two orders of magnitude ($99\%$). Moreover, assessment of the image quality confirms that the integrity of the image displayed is largely preserved. In fact, only a minor fraction of the image-forming light is diverted to the corrective beam. This is approximately 7$\%$, calculated as the relative intensity reduction in the point-cloud hologram region excluding the ZOD spot between uncorrected and corrected images. The peak-to-valley (PV) uniformity ($U_{PV}$) has been quantified as:
\begin{equation}
U_{PV} = \frac{I_{min}}{I_{max}}
\end{equation}
where $I_{min}$ and $I_{max}$ denote the minimum and maximum pixel intensities within the reference points of the point-cloud image. This metric ranges from 0 to 1, with unity corresponding to perfectly uniform intensity distribution. In our experiments, the $U_{PV}$ of the reconstructed point cloud decreased only marginally from 0.85 to 0.83 after correction, indicating that the near-complete ZOD suppression had a negligible impact on the perceived uniformity of the holographic image.

To test the robustness and generality of our method, the corrective phase map derived from the point-cloud experiment was then applied to a more complex, 2D layer-based hologram of a tree as shown in Fig. \ref{fig:holograms} from (a) to (f). While the corrective phase pattern remains consistent for different holograms, the amplitude weights for the image and the corrective field were adjusted to match the specific ZOD intensity of this new hologram. Even with a different and more complex image, the ZOD was effectively suppressed to a residual normalized intensity of $0.026$. In this case, since the ZOD overlaps with the image itself, the goal of the correction is not a complete elimination of the ZOD light, but rather a suppression consistent with the average intensity level of the image.  Crucially, the correction did not degrade the holographic content, as confirmed by the Peak Signal-to-Noise Ratio (PSNR), Mean Square Error (MSE) and structural similarity index values reported in \ref{fig:holograms}(d) and \ref{fig:holograms}(e) which remained almost unchanged (PSNR shifted from 8.1 dB to 8 dB), confirming the stability of the corrective phase map.

\begin{figure*}[htp]
\centering
\includegraphics[width=\textwidth]{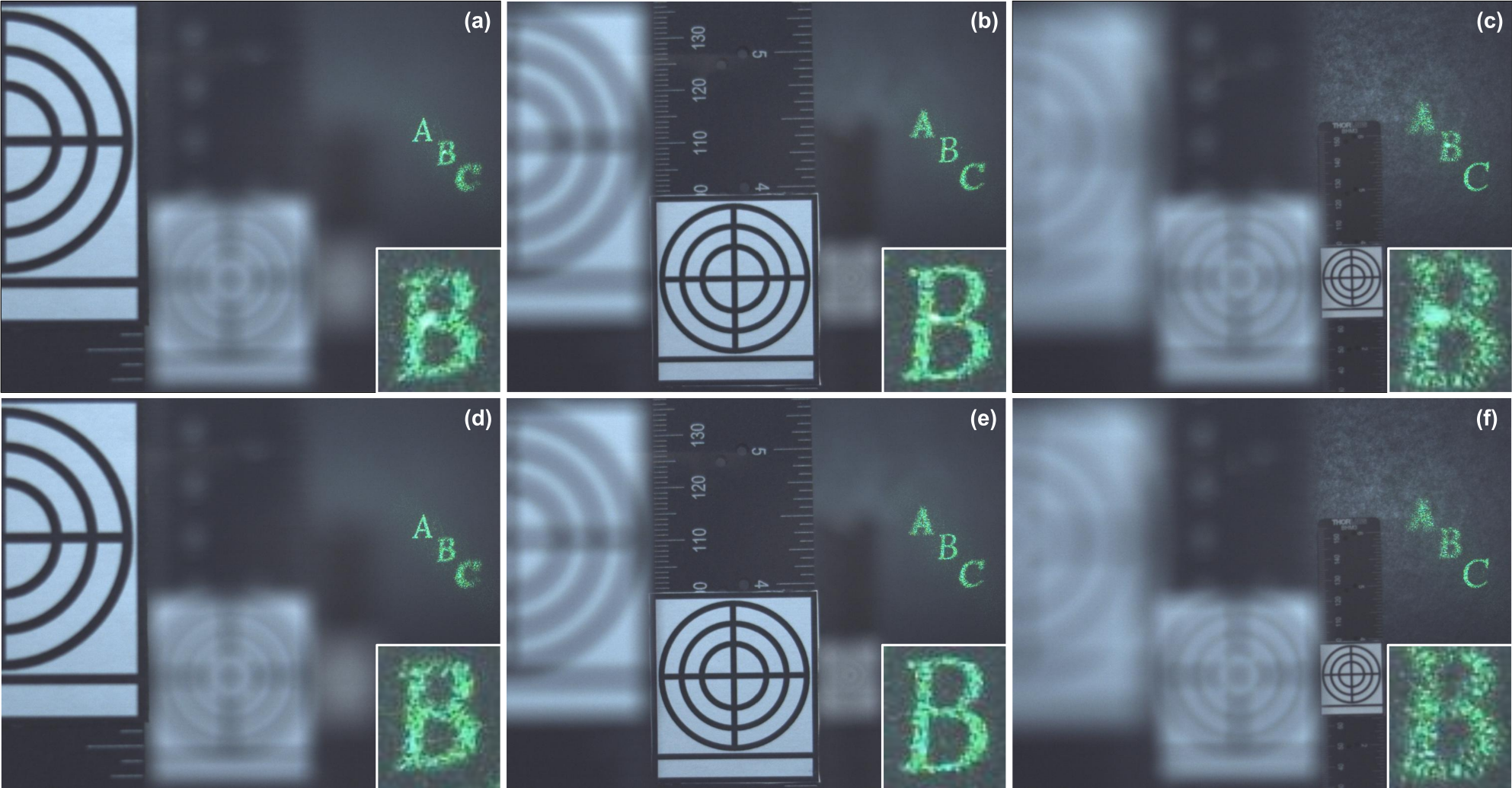}
\caption{Qualitative assessment of ZOD suppression in multi-plane holographic projections within a mixed-reality environment. Panels (a)–(c) illustrate uncorrected holograms exhibiting haze and reduced contrast across focal planes: (a) closest to the observer, (b) intermediate distance, and (c) farthest from the observer. Panels (d)–(f) show corresponding corrected projections with ZOD suppression, yielding sharper, high-contrast images without central artifacts. Insets provide magnified views of ZOD region highlighting enhanced fidelity. The setup confirms robust 3D rendering for augmented reality, with ZOD elimination essential for artifact-free multi-depth visualization.}
\label{fig:AR}
\end{figure*}

In Fig. \ref{fig:AR}, we qualitatively illustrate the method’s performance in a realistic AR NED scenario. For this demonstration, we emphasize visual assessment, since precise intensity measurements in out-of-focus, multi-plane scenes are experimentally impractical. The uncorrected holograms (Fig. \ref{fig:AR} from (a) to (c)) display the typical artifacts of unsuppressed ZOD, with a pervasive intensity component concentrated at the center of the reconstructed field. In contrast, the corrected holograms (Fig. \ref{fig:AR} from (d) to (f)) show the same multi-plane projection with ZOD suppression applied. The results show significant ZOD reduction and improved reconstruction of the original image. Magnified insets further emphasize the gain in fidelity. These results demonstrate the effectiveness of the technique in real 3D projection and highlight ZOD elimination as a critical requirement for artifact-free multi-depth visualization in immersive AR systems.

The input beam on the SLM was set to be slightly diverging. This alignment provides specific advantages for augmented-reality applications, particularly in the “arm’s reach” regime, where the goal is to generate virtual objects at close distances. In this configuration, the Fourier plane is brought to a finite distance, enabling the rendering of images closer to the observer. In our setup, the Fourier plane, where the zero-order diffraction becomes more pronounced as it comes into focus, coincides with the position marked by the target letter B. Fig. \ref{fig:AR} illustrates that the closed-loop ZOD removal is agnostic to such alignment variations: while the suppression is designed and validated at the Fourier plane, its benefits extend throughout the entire depth of field, ensuring artifact-free 3D reconstructions.

A variety of example datasets, generated either as Fourier-transform based holograms or as wireframe holograms \cite{Astarita:25} are available as a public repository \cite{pozzi_2026_18254155}.

\section{Discussion}
The removal of ZOD through a destructive interference approach is of significant relevance for several reasons. Firstly, the retrieved ideal phase map remains consistent across different holograms, ensuring conservation of the full modulation depth of the SLM. Secondly, the computation of the phase pattern is both straightforward and efficient, facilitating rapid implementation in practical applications. Additionally, the interfering system is always inherently aligned since both beams originate from the same display and follow the same optical path, making it a robust approach to achieve ZOD removal.

However, the intensity of the ZOD can slightly depend on the specific holographic patterns, which requires the estimation of coefficients $A_1$ and $A_2$ in advance. This pre-estimation process enhances the robustness and predictability of the approach, ultimately improving its integration for actual applications.

Moreover, because the corrective replica draws only a small fraction of the modulated light, the overall diffraction efficiency is only marginally reduced while the image benefits from the reduced on-axis artefacts. Consistently with the experimental validation in Fig.~\ref{fig:AR}, the residual intensity loss does not compromise hologram visibility under standard ambient illumination in our augmented-reality demonstrator.

It should also be stated that, after the compensation phase pattern is measured, compensation of any hologram requires negligible computation time. We estimated the compensation time for the reported setup in less than $5\, ms$ when computing on a CPU, and down to the $\mu s$ scale when the compensation is integrated in the GPU computation of holograms required for real time display applications.

Our understanding is that this method maximizes the interferometric suppression efficiency allowed by the experimental setup, and residual zeroth order light after compensation is only due to imperfect polarization of the laser, the limited fill factor and resolution of the SLM, and the phase flickering due to the digital nature of the pixel electronics \cite{yang2020phase}.
The proposed method can be adapted to a wide variety of spatial phase modulator devices, such as transmissive liquid crystal devices or microelectromechanical devices. The only strong requirement is that the diffraction efficiency of the device (and therefore its fill factor) greatly exceeds $50 \%$, as required by equation \ref{eq:A1}. Unfortunately, due to the radically different hologram generation principle, the described method does not apply to intensity modulation SLMs.

While the calculations and demonstrations in this study focus on monochromatic holograms, the method can be trivially generalized to standard multicolor hologram projection. While some alternative methods exist \cite{kavakli2023multi}, state-of-the-art full-color holograms are predominantly achieved through time multiplexing, which enables dynamic content without compromising spatial resolution and is favored over spatial multiplexing approaches (better suited for static scenes) due to reduced crosstalk and higher efficiency \cite{gopakumar2024fullcolour, Yoshida:24}. Our method can readily be extended to color setups by applying wavelength-specific corrections for the three primary colors (R, G, B), effectively retrieving three independent RGB phase maps via the CITL procedure, a straightforward adaptation that leverages the technique's compatibility with any CGH algorithm. As next steps, we plan to implement this RGB extension using the Holoeye GAEA device, leveraging the predominant time-multiplexing approach for full-color holograms to ensure seamless integration with dynamic AR content while maintaining spatial resolution.

\section{Conclusions}

This manuscript has presented what we are convinced represents the best available compromise for the suppression of zeroth order artifacts in phase-only CGH.
By deepening our understanding and control of destructive interference, this approach significantly enhances the clarity and fidelity of holographic images, while preserving the full field of view.
Beyond its experimental validation, this approach represents a conceptual shift in holographic beam control: zero-order suppression is achieved directly within the phase encoding process, eliminating the need for external filtering optics. By embedding interference engineering into the CGH computation, the method bridges a long-standing gap between laboratory holography and compact display integration.
As this technique does not add any complexity to the setup, it is a significant advancement for applications such as augmented reality near-eye displays, expanding the possibilities of holographic display technology.
The ability to perform full-field ZOD cancellation in a compact, alignment-free configuration marks a step toward deployable holographic engines for near-eye and wearable displays. More broadly, the same camera-in-the-loop interferometric principle could be even adapted to compensate optical aberrations, across a wide range of diffractive systems.


\section*{Acknowledgments}
This work was carried out in the Smart Eyewear Lab, a Joint Research Platform between EssilorLuxottica and Politecnico di Milano.

\section{Disclosures}
This work was funded by EssilorLuxottica S.A. and a patent application related to the presented research has been submitted

\section{Data availability}
Data underlying the results presented in this paper are available in Ref.  \cite{pozzi_2026_18254155}.

\nolinenumbers

\bibliography{bibliografia}

@article{piccardo2023broadband,
  title={Broadband control of topological--spectral correlations in space--time beams},
  author={Piccardo, Marco and de Oliveira, Michael and Policht, Veronica R and Russo, Mattia and Ardini, Benedetto and Corti, Matteo and Valentini, Gianluca and Vieira, Jorge and Manzoni, Cristian and Cerullo, Giulio and others},
  journal={Nature Photonics},
  volume={17},
  number={9},
  pages={822--828},
  year={2023},
  publisher={Nature Publishing Group UK London}
}

@article{yang2020phase,
  title={Phase flicker in liquid crystal on silicon devices},
  author={Yang, Haining and Chu, DP},
  journal={Journal of Physics: Photonics},
  volume={2},
  number={3},
  pages={032001},
  year={2020},
  publisher={IOP Publishing}
}

@article{pozzi2018fast,
  author = {Paolo Pozzi and Laura Maddalena and Nicolò Ceffa and Oleg Soloviev and Gleb Vdovin and Elizabeth Carroll and Michel Verhaegen},
  title = {Fast calculation of computer generated holograms for 3D photostimulation through compressive-sensing Gerchberg–Saxton algorithm},
  journal = {Methods and Protocols},
  year = {2018},
  volume = {2},
  issue = {1},
  pages = {2},
  publisher = {MDPI},
  doi = {10.3390/mps2010002}
}

@dataset{pozzi_2026_18254155,
  author       = {Pozzi, Paolo},
  title        = {Additional example images for the manuscript
                   "Zero-Order Diffraction Suppression in Full Field-
                   of-View Computer Generated Holography: A Camera In
                   the Loop Interferometric Approach"
                  },
  month        = jan,
  year         = 2026,
  publisher    = {Zenodo},
  version      = {Pre-publication version},
  doi          = {10.5281/zenodo.18254155},
  url          = {https://doi.org/10.5281/zenodo.18254155},
}

@article{grier2003revolution,
  author = {David Grier},
  title = {A revolution in optical manipulation},
  journal = {Nature},
  year = {2003},
  volume = {424},
  pages = {810--816},
  doi = {10.1038/nature01935}
}

@inproceedings{kavakli2023multi,
  title={Multi-color holograms improve brightness in holographic displays},
  author={Kavakl{\i}, Koray and Shi, Liang and Urey, Hakan and Matusik, Wojciech and Ak{\c{s}}it, Kaan},
  booktitle={SIGGRAPH Asia 2023 Conference Papers},
  pages={1--11},
  year={2023}
}

@article{gopakumar2024fullcolour,
  title={Full-colour 3D holographic augmented-reality displays with metasurface waveguides},
  author={Manu Gopakumar and Gun-Yeal Lee and Suyeon Choi and Brian Chao and Yifan Peng and Jonghyun Kim and Gordon Wetzstein},
  journal={Nature},
  volume={629},
  pages={791--797},
  year={2024},
  doi={10.1038/s41586-024-07386-0}
}

@article{jang2024waveguide,
  title={Waveguide holography for 3D augmented reality glasses},
  author={Changwon Jang and Kiseung Bang and Minseok Chae and Byoungho Lee and Douglas Lanman},
  journal={Nature Communications},
  volume={15},
  pages={66},
  year={2024},
  doi={10.1038/s41467-023-44032-1}
}

@inproceedings{Improso2017SuppressionOZ,
  title={Suppression of Zeroth-Order Diffraction in Phase-Only Spatial Light Modulator via Destructive Interference with a Correction Beam},
  author={Wynn Dunn Gil Dugang Improso and Giovanni Tapang and Caesar A. Saloma},
  booktitle={International Conference on Photonics, Optics and Laser Technology},
  year={2017},
  url={https://api.semanticscholar.org/CorpusID:53317741}
}

@article{palima2007holographic,
  title={Holographic projection of arbitrary light patterns with a suppressed zero-order beam},
  author={Palima and Darwin and Daria and Vincent Ricardo},
  journal={Applied Optics},
  volume={46},
  number={20},
  pages={4197--4201},
  year={2007},
  publisher={Optical Society of America},
  doi={10.1364/AO.46.004197}
}

@article{pan2015review,
  title={A review of dynamic holographic 3D display: algorithms, devices, and systems},
  author={Yijie Pan and Juan Liu, Xin Li and Yongtian Wang},
  journal={IEEE Transactions on Industrial Informatics},
  volume={12},
  pages={1--1},
  year={2015},
  doi={10.1109/TII.2015.2496304}
}

@article{ronzitti2012LCOS,
  title={LCoS nematic SLM characterization and modeling for diffraction efficiency optimization, zero and ghost orders suppression},
  author={Emiliano Ronzitti, Marc Guillon, Vincent de Sars and Valentina Emiliani},
  journal={Optics Express},
  volume={20},
  number={16},
  pages={17843--17855},
  year={2012},
  publisher={Optical Society of America}
}

@article{CGHalg2022,
  author    = {Dapu Pi and Juan Liu and Yongtian Wang},
  title     = {Review of computer-generated hologram algorithms for color dynamic holographic three-dimensional display},
  journal   = {Light: Science and Applications},
  year      = {2022},
  volume    = {11},
  pages     = {231},
  doi       = {10.1038/s41377-022-00916-3},
  url       = {https://doi.org/10.1038/s41377-022-00916-3}
}

@article{liang2012ZOD,
  title = {Suppression of the zero-order diffracted beam from a pixelated spatial light modulator by phase compression},
  author = {Jinyang Liang and Sih-Ying Wu and Fredrik K. Fatemi and Michael F. Becker},
  journal = {Applied Optics},
  volume = {51},
  number = {16},
  pages = {3294--3304},
  year = {2012},
  publisher = {Optical Society of America},
  doi = {10.1364/AO.51.003294},
  url = {https://doi.org/10.1364/AO.51.003294}
}

@article{hernandez2014ZOD,
  title = {Zero-order suppression for two-photon holographic excitation},
  author = {Oscar Hernandez and Marc Guillon and Eirini Papagiakoumou and Valentina Emiliani},
  journal = {Optics Letters},
  volume = {39},
  number = {20},
  pages = {5953--5956},
  year = {2014},
  publisher = {Optical Society of America},
  doi = {10.1364/OL.39.005953},
  url = {https://doi.org/10.1364/OL.39.005953}
}

@article{yu2024performance,
author = {Yu, Mr and Ren, Suxia and Xuan, Hongwen and Wang, Yisen},
year = {2024},
month = {07},
pages = {29483-29493},
title = {Performance improved for two-photon multi-focus microscopy based on a spatial light modulator by eliminating the zero-order beam},
volume = {32},
journal = {Optics Express},
doi = {10.1364/OE.532120}
}

@article{Astarita:25,
author = {Marco Astarita and Alessandro Cerioni and Andrea Bassi and Matteo Ziliani and Anna Cesaratto and Tommaso Ongarello and Giulio Cerullo and Gianluca Valentini and Paolo Pozzi},
journal = {Opt. Express},
number = {14},
pages = {30162--30173},
publisher = {Optica Publishing Group},
title = {Wireframe holography as a method for augmented reality projections},
volume = {33},
month = {Jul},
year = {2025},
url = {https://opg.optica.org/oe/abstract.cfm?URI=oe-33-14-30162},
doi = {10.1364/OE.561735},
}

@article{Yoshida:24,
author = {Shuhei Yoshida},
journal = {Appl. Opt.},
number = {10},
pages = {2455--2461},
publisher = {Optica Publishing Group},
title = {High-speed full-color computer-generated holography using a digital micromirror device and fiber-coupled RGB laser diode},
volume = {63},
month = {Apr},
year = {2024},
url = {https://opg.optica.org/ao/abstract.cfm?URI=ao-63-10-2455},
doi = {10.1364/AO.509556},
}

@article{VAC,
  author = {Kramida, Gregory},
  title = {Resolving the Vergence-Accommodation Conflict in Head-Mounted Displays},
  journal = {IEEE Transactions on Visualization and Computer Graphics},
  volume = {22},
  number = {7},
  pages = {1912--1931},
  year = {2016},
  doi = {10.1109/TVCG.2015.2473855}
}

@article{Yun:23,
author = {Xue Yun and Yansheng Liang and Minru He and Linquan Guo and Xinyu Zhang and Tianyu Zhao and Piero R. Bianco and Ming Lei},
journal = {Opt. Express},
number = {12},
pages = {19613--19621},
publisher = {Optica Publishing Group},
title = {Zero-order free holographic optical tweezers},
volume = {31},
month = {Jun},
year = {2023},
url = {https://opg.optica.org/oe/abstract.cfm?URI=oe-31-12-19613},
doi = {10.1364/OE.489014},
}

@article{LIANG2022107048,
title = {Zero-order-free complex beam shaping},
journal = {Optics and Lasers in Engineering},
volume = {155},
pages = {107048},
year = {2022},
issn = {0143-8166},
doi = {https://doi.org/10.1016/j.optlaseng.2022.107048},
url = {https://www.sciencedirect.com/science/article/pii/S0143816622001038},
author = {Yansheng Liang and Xue Yun and Minru He and Zhaojun Wang and Shaowei Wang and Ming Lei}
}

\bibliographystyle{abbrv}

\end{document}